# Ensemble Kalman filter meets model predictive control in chaotic systems

Yohei Sawada[1]

[1] Department of Civil Engineering, Graduate School of Engineering, the University of Tokyo, Tokyo, Japan

Corresponding author: Y. Sawada, Department of Civil Engineering, the University of Tokyo, Tokyo, Japan, 7-3-1, Hongo, Bunkyo-ku, Tokyo, Japan, yoheisawada@g.ecc.u-tokyo.ac.jp

**Abstract**

Although data assimilation originates from control theory, the relationship between modern data assimilation methods in geoscience and model predictive control has not been extensively explored. In the present paper, I discuss that the modern data assimilation methods in geoscience and model predictive control essentially minimize the similar quadratic cost functions. Inspired by this similarity, I propose a new ensemble Kalman filter (EnKF)-based method for controlling spatio-temporally chaotic systems, which can be applied to high-dimensional and nonlinear Earth systems. In this method, the reference vector, which serves as the control target, is assimilated into the state space as a pseudo-observation by ensemble Kalman smoother to obtain the appropriate perturbation to be added to a system. A proof-of-concept experiment using the Lorenz 63 model is presented. The system is constrained in one wing of the butterfly attractor without tipping to the other side by reasonably small control perturbations which are comparable with previous works.

# 1. Introduction

Data assimilation in geoscience is a fundamental technique to monitor, predict, and understand Earth systems. Originating from control theory (Kalman, 1960), data assimilation in geoscience has uniquely evolved from control theoretical state estimation methods to efficiently estimate the state variables of extremely high-dimensional and nonlinear Earth systems from sparsely distributed observations. The relationship between modern data assimilation methods in geoscience and control theory has not recently been explored.

However, data assimilation in geoscience and model predictive control share a strong connection. The four-dimensional variational method (4D-Var), a widely used data assimilation method in numerical weather prediction, essentially aims to minimize the following cost function (e.g., Talagrand 2014; Bannister 2017),

$$J(x_0) = \frac{1}{2}(x_0 - x_0^b)^T B^{-1}(x_0 - x_0^b) + \sum_{t=0}^{T} \frac{1}{2}(y_t - H(x_t))^T R^{-1}(y_t - H(x_t)) \quad (1)$$

$$s.t. \; x_{t+1} = M(x_t)$$

where $x_t$ is the state variables at time t, $x_0^b$ is the background of $x_0$, $B$ is the background error covariance matrix, $y_t$ is the observations at time $t$, $R$ is the observation error covariance matrix, $H$ is the observation operator, $T$ is the data assimilation window length, and $M$ is a model which describes the temporal evolution of the state variables. Note that $H$ and $R$ are assumed time-invariant in Equation (1). The first term on the right-hand side of Equation (1) increases when estimated state variables deviates from the initial guess (i.e. background), while the second term increases when model-predicted observable variables within the assimilation window deviate from observation. Model predictive control is a broad class of control methods which use process-based models to control the future behavior of the controlled system (e.g., Schwenzer et al. 2021 for a comprehensive review). The goal of model predictive control is to find the smallest control inputs to minimize the difference between future states and a control objective. Although the cost function minimized in model predictive control varies by problem setting, a typical quadratic cost function can be expressed as:

$$J(u_0, u_1, \ldots, u_{T_c-1}) = \sum_{t=0}^{T_c-1} u_t^T C^{u-1} u_t + \sum_{t=1}^{T_c} (r_t - H^c(x_t))^T C^{r-1}(r_t - H^c(x_t)) \quad (2)$$

$$s.t. \; x_{t+1} = M(x_t, u_t)$$

where $u_t$ is control inputs at time $t$, $T_c$ is the control horizon, $r_t$ is the reference vector indicating the desired state at time $t$, $H^c$ is the operator to map state variables onto control variables for comparison with the reference variables, and $C^u$ and $C^r$ are user-defined weights. The first term on the right-hand side of Equation (2) increases when external control inputs become large, while the second term increases when model-predicted states deviate from the control objective. Minimizing equation (2) finds the smallest control inputs or perturbations needed to reasonably minimize the difference between the future state and the control objective, based on model predictions. When $u_0$ is defined as $x_0 - x_0^b$ of Equation (1), the similarity between 4D-Var in geoscience and model predictive control becomes evident. Minimizing Equation (1) seeks the smallest perturbation to be added to the initial state (i.e. background) to effectively minimize the difference between model-predicted observations and actual observations.

Henderson et al. (2005) drew upon this analogy between 4D-Var in numerical prediction systems and model predictive control to conduct a numerical experiment aimed at mitigating a tropical cyclone. They modified the second term on the right-hand side of Equation (1) as follows:

$$J(x_0) = \frac{1}{2}(x_0 - x_0^b)^T B^{-1}(x_0 - x_0^b) + \lambda \sum_{t=0}^{T_c} J_d(x_t) \quad (3)$$

$$s.t. \, x_{t+1} = M(x_t)$$

where $J_d$ represents the damage function in which the economic damage caused by a tropical cyclone is parameterized by wind speed at coastal cities, and $\lambda$ is a weighting factor. Minimizing Equation (3) seeks the smallest perturbation to be added to the initial state to effectively minimize the predicted economic damage. By leveraging the existing framework of 4D-Var in an atmospheric model, Henderson et al. (2005) demonstrated that the appropriate perturbation to mitigate damages from a tropical cyclone can be estimated using this 4D-Var-based control approach. Despite a lot of ideas proposed to modify the chaotic and extreme weather events (e.g., Wiloughby et al. 1985; Cotton et al. 2007; Breed et al. 2013; Latham et al. 2012; Jacobson et al. 2014), the efficient estimation of optimal perturbations to the atmosphere has not been extensively investigated since the pioneering work of Henderson et al. (2005). Specifically, the potential of ensemble data assimilation methods, such as an ensemble Kalman filter (EnKF; see e.g., Houtekamer and Zhang 2016), has not been explored in the context of model predictive control.

Here I present a method to control spatio-temporal chaotic systems using EnKF. The proposed method is deeply influenced by the similarity between the modern data assimilation methods in geoscience and model predictive control, as discussed above. The core idea is to solve the minimization of Equation (2), which is the problem solved in model predictive control, using EnKF. A significant advantage of my proposed method is that it directly leverages the existing EnKF architecture, which has been widely shown effective and flexible for geoscientific applications. I show EnKF, with its iteration-free and derivative-free nature, makes control problems in geoscience easy to solve. Miyoshi and Sun (2022) proposed controlling a chaotic system through ensemble prediction. Although they used EnKF for state estimation, their control method does not directly employ EnKF and does not fully use the information of ensemble, which differs from what I propose in this paper. Kawasaki and Kotsuki (2024) also applied EnKF for estimating current states. They adopted a conventional model predictive control method used in control engineering, which is computationally expensive and again differs from what I propose in this paper since I fully rely on EnKF to solve model predictive control problem and estimate the perturbations required to control a system. Note also that I intend to propose a method suitable for controlling systems with extremely large degree of freedom, in which the size of state vectors is the order of $10^4 \sim 10^9$, such as atmosphere, while most methods in control engineering are fine tuned to control smaller size systems.

## 2. Method

Following the approach of Henderson et al. (2005), I propose a control method by modifying the cost function minimized by EnKF. In a filtering scenario, where the analysis time coincides with the observation time, the EnKF aims to minimize the following cost function:

$$J(x_0) = \frac{1}{2}(x_0 - \overline{x_0^b})^T {P^b}^{-1}(x_0 - \overline{x_0^b}) + \frac{1}{2}(y_0 - H(x_0))^T R^{-1}(y_0 - H(x_0)) \qquad (4)$$

where $\overline{x_0^b}$ is the background ensemble mean of state estimates from the ensemble, $P^b$ is the background error covariance matrix estimated from ensemble members. Assuming that the observation operator is linear, and errors follow the Gaussian distribution, EnKF solution minimizes Equation (4). It is not necessary to obtain a full covariance matrix $P^b$ as well as the linearized observation operator since ensemble-based approximations are used to compute Kalman gain (see equation (1)-(10) in Houtekamer and Zhang 2016). There are several flavors of EnKF to transport each ensemble members. In this paper, I used the ensemble transform Kalman filter (ETKF; Bishop et al. 2001, Hunt et al. 2007) to obtain the analysis ensemble. ETKF transports a background ensemble $\{x_0^{b(i)}: i = 1,2,......,k\}$ to an analysis ensemble $\{x_0^{a(i)}: i = 1,2,......,k\}$ ($k$ is the ensemble size) using the following equations:

$$X^a = X^b W^a \quad (5)$$

$$W^a = [(k-1)\tilde{P}^a]^{\frac{1}{2}} \quad (6)$$

$$\tilde{P}^a = [(k-1)I + (Y^b)^T R^{-1} Y^b]^{-1} \quad (7)$$

where $X^b$ and $X^a$ includes perturbations of state variables of background and analysis ensemble members, respectively. The $i$th column of $X^b$ is $x_0^{b(i)} - \overline{x_0^b}$. $Y^b$ is analogous to $X^b$, and the $i$th column of $Y^b$ is $y_0^{b(i)} - \overline{y_0^b}$ where $y_0^{b(i)} = H(x_0^{b(i)})$. Note that the proposed control algorithm can be applied to the other flavors of EnKF.

After obtaining the analysis ensemble, an extended ensemble forecast is performed from the analysis ensemble members over the control horizon, $T_c$. Then, another minimization problem is solved using the following quadratic cost function, which aligns with typical model predictive control practices and the approach of Henderson et al. (2005):

$$J^c(x_0) = \frac{1}{2}(x_0 - \overline{x_0^a})^T C^{u-1}(x_0 - \overline{x_0^a}) + \sum_{t=0}^{T_c} \frac{1}{2}(r_t - H^c(x_t))^T C^{r-1}(r_t - H^c(x_t)) \quad (8)$$

$$s.t. x_{t+1} = M(x_t)$$

where $\overline{x_0^a}$ is the analysis ensemble mean. In the proposed method, $C^u$ is set to $P^a$ which is the analysis error covariance matrix. In addition, the predicted system is evaluated only at the end of the control horizon in this paper. Therefore, the cost function for system control in this paper can be expressed as:

$$J^c(x_0) = \frac{1}{2}(x_0 - \overline{x_0^a})^T P^{a-1}(x_0 - \overline{x_0^a}) + \frac{1}{2}(r_{T_c} - H^c(x_{T_c}))^T C^{r-1}(r_{T_c} - H^c(x_{T_c})) \quad (9)$$

It is straightforward to recognize this minimization problem as ensemble Kalman smoother (EnKS; e.g., Cosme, 2014; Evensen and van Leeuwen 2000). By setting $C^u$ to $P^a$, the ensemble estimated by EnKF can directly be used. In addition, it is expected to effectively use the information of correlations between state variables to estimate effective interventions to achieve a control objective. To solve this control problem, first, the model-based ensemble prediction is projected onto the control criteria. Then, the minimization of Equation (9) can be achieved by assimilating $r_{T_c}$ as a "pseudo-observation" with the "pseudo-observation error covariance", $C^r$, into the analysis state variables using ETKF. Through this process, the appropriate perturbation to be added to $\overline{x_0^a}$ can be obtained as an analysis increment. The proposed EnKF-based control algorithm, called Ensemble Kalman Control (EnKC), is outlined in Algorithm 1.

## 3. Experiment design

The proof-of-concept numerical simulation in this paper is consistent to Miyoshi and Sun (2022). The Lorenz 63 model (Lorenz 1963) was used to test the proposed algorithm:

$$\frac{dX}{dt} = -10(X - Y) \tag{10}$$

$$\frac{dY}{dt} = -XZ + 28X - Y \tag{11}$$

$$\frac{dz}{dt} = XY - \frac{8}{3}Z \tag{12}$$

This model was numerically solved using the fourth-order Runge-Kutta method with the timestep of 0.01. The time interval between observations was set to 8 timesteps. Observation error (standard deviation) was set to $\sqrt{2}$, and it was assumed that all variables $X$, $Y$, and $Z$ were observed. The observation errors for all three variables are uncorrelated. Observations were generated from the nature run by adding Gaussian noises. A total of 128000 timesteps (=16000 data assimilation cycles) were performed. The first 2500 timesteps were discarded as spin-up. The results in all timesteps but the spin-up period were used for evaluation. Following Miyoshi and Sun (2022), the initial condition of the nature was chosen to be $(X, Y, Z) = (8.20747, 10.0860, 23.8632)$, which is aligned with the attractor of Equations (10)-(12). The initial conditions of ensemble members were generated by adding Gaussian white noises, whose mean is 0 and standard deviation is $\sqrt{2}$, to this nature's initial condition. The ensemble size was set to 3. I fully followed the framework of a control simulation experiment as proposed by Miyoshi and Sun (2022). The state variables of the nature run were altered by control measures (specifically, in Step 3 of Algorithm 1). Consequently, I sequentially simulated the nature and generated observations at each data assimilation step.

Same as Miyoshi and Sun (2022), the control objective is staying in a wing of the butterfly attractor where $X$ is positive, without tipping to the other side. To achieve this, I defined our control operator, $H^c$, as a logistic function:

$$H^c(x) = \frac{1}{1 + \exp(-X)} \tag{13}$$

When $X \gg 0$, $H^c(x)$ goes to 1. Thus, it is logical to set $r_{T_c}$ in Equation (9) to 1.0. At every control step, model-predicted states were evaluated by Equation (13), and the pseudo-observation $r_{T_c} = 1.0$ was assimilated into the analyzed state space to obtain the appropriate perturbation. Since the control criterion is a scalar (i.e. the reference vector is actually a scaler), $\boldsymbol{C}^r$ in this paper is also a scaler. I varied $\boldsymbol{C}^r$ to $10^{-1}$, $10^{-2}$, $10^{-3}$, $10^{-4}$, $10^{-5}$ and $10^{-6}$ to examine its sensitivity to the performance. The control horizon, $T_c$, was set to 10, 50, 100, and 300 timesteps. Note that the control objective of $X > 0$ is almost identical to that of $Y > 0$, so that the similar results can be obtained by using $Y$ instead of $X$ in Equation (13).

I also performed the control method proposed by Miyoshi and Sun (2022) to illustrate the characteristics of EnKC. In Miyoshi and Sun (2022), EnKF is performed and then the extended forecast is provided during the control horizon, $T_c$. To determine the direction of the intervention, two ensemble members of the extended forecast are used. One is a member showing the regime shift and another is a member not showing the regime shift. The difference of the state variables between these two ensemble members is used as a vector showing the intervention's direction. If no member is in the desired regime, the ensemble members from the former initial time are used to perform the extended forecast. The magnitude of the control perturbations is scaled to a prescribed norm, $D_{fix}$. In this paper, all three variables are perturbed. I found that it was necessary to perform covariance inflation to successfully implement the method of Miyoshi and Sun (2022) while the inflation was unnecessary for EnKC. This is probably because Miyoshi and Sun (2022) needs to effectively sample both tipping and non-tipping members from extended forecast, so that their ensemble spread should be kept large. I performed the relaxation to prior perturbation (RTPP) method (Zhang et al. 2004) with $\alpha = 0.9$ for Miyoshi and Sun (2022). See Algorithm 2 for the implementation of Miyoshi and Sun (2022)

## 4. Results

Figures 1a and 1b show the attractor of the uncontrolled Lorenz 63 nature run and the controlled Lorenz 63 nature run with $\boldsymbol{C}^r = 10^{-6}$ and $T_c = 300$, respectively. Figures 1c and 1d clearly demonstrate that the nature run can be controlled to stay in one wing of the butterfly attractor if an appropriate weight is selected. Figure 2a shows the typical timeseries of $X$, indicating the controlled system stays in a periodic orbit. Figure 2b shows the magnitude of the perturbation defined as $D = \|x_0 - \overline{x_0^a}\|$ (see also Step 3 described in Algorithm 1 for more details). The perturbation is significantly smaller than state variables.

The sensitivity of $\boldsymbol{C}^r$ (its inverse is the user-defined weight for control criteria) and $T_c$ (control horizon) to the performance is discussed. Figures 3a-d indicate that, to successfully control the system, it is necessary to choose a small $\boldsymbol{C}^r$, which assigns a larger weight to meeting the control criterion over minimizing the perturbation. Figures 3e-h show the magnitude of the perturbation. Note that a logarithm scale is used in Figures 3e-h. As $\boldsymbol{C}^r$ decreases, the minimum magnitude of the perturbation added to the system increases. When more weights are given to the second term on the right-hand side of Equation (6) to meet the control criterion, a larger external control force is required from ETKF. On the other hand, the maximum magnitude of the added perturbation does not greatly change as $\boldsymbol{C}^r$ varies. When $\boldsymbol{C}^r$ is too large to effectively control the system, large deviations from the idealized condition often arise, which requires to generate large perturbations. Therefore, large perturbations can be produced even with small weights assigned to the second term on the right-hand side of Equation (6). When a sufficiently small $\boldsymbol{C}^r$ is chosen, EnKC with relatively shorter control horizons can successfully achieve the control objective. While the trajectory stays in the smaller region ($6.5 < X < 10.5$) when $T_c$ is 300 timesteps (see also Figure 2a), it stays in larger regions with shorter $T_c$. This may be an advantage of choosing shorter control horizons if the substantial modification of the trajectory is recognized as the adverse effect of the control. However, the magnitude of the perturbation, $D$, increases when $T_c$ becomes shorter. This result clearly indicates that the effect of small perturbation added at the beginning of the control horizon can be amplified to efficiently control the system if the control horizon is long. When the control horizon is short, the growth of the perturbation during the control horizon cannot be effectively leveraged. However, it should be noted that the system is found to leave the original attractor when $\boldsymbol{C}^r = 10^{-4}$ and $T_c = 300$ (Figure 3d). I found that the long control horizons and large weights sometimes make the estimation of control perturbation unstable when the system stays in the undesired state (i.e. $X < 0$). This point should be noted toward real-world applications.

The control method proposed by Miyoshi and Sun (2022) was also performed and compared with EnKC. I set $D_{fix} = 0.05$ and $T_c = 300$ since they are optimal (see Figure 3 of Miyoshi and Sun (2022)). The rightmost violin plot of Figure 4a shows that the control by Miyoshi and Sun (2022) is not always successful even under the optimal setting, which is consistent with the original paper. The control of EnKC is always successful during the 16000 assimilation cycles. It should also be noted that EnKC works with the shorter control horizons while Miyoshi and Sun (2022) indicated that their method does not work with $T_c < 200$. Figure 4b compares the mean magnitude of control perturbations during the assimilation window (8 timesteps). As discussed in the previous paragraph, the magnitude of control perturbations becomes larger when shorter control horizons are chosen in the proposed method. Under the same control horizon (i.e., $T_c = 300$), the magnitude of control perturbations required in the proposed method is smaller than Miyoshi and Sun (2022). Overall, EnKC outperforms Miyoshi and Sun (2022) in terms of both control accuracy and cost. The advantage of EnKC against Miyoshi and Sun (2022) is that the information of ensemble forecast can be fully exploited using an ensemble covariance matrix in EnKS while Miyoshi and Sun (2022) used only the differences between two representative ensemble members.

**5. Discussion and Conclusions**

In this paper, I presented a method to control spatio-temporal chaotic systems. This proposed method is inspired by the similarity between the modern data assimilation methods in geoscience and model predictive control. A significant advantage of our proposed method, EnKC, is the ability to directly employ the existing EnKF architecture, which has been widely used for state estimation of Earth systems, for control purposes. The EnKF, with its iteration-free and derivative-free nature, offers distinct benefits over other data assimilation methods, such as 4D-Var. This advantage is leveraged in EnKC. Additionally, recent advancements in EnKF within geoscience, such as iterative smoothers for addressing nonlinearity of the system (e.g., Bocquet and Sakov 2012), localization methods to mitigate sampling errors (e.g., Hunt et al. 2007), inflating observation error covariances to prevent from providing too large increments (e.g., Minamide and Zhang 2017), flow-dependent uncertainty quantification to account for model imperfectness (e.g., Sawada and Duc 2024), can seamlessly be integrated into the proposed control method. Consequently, the proposed EnKF-based control method is particularly well-suited for application in Earth system sciences, in which EnKF is already well established, including weather modification. Although it can be used to control any spatio-temporal chaotic systems, its potential to control high-dimensional systems should be thoroughly evaluated using a real-world case as a future work.

Miyoshi and Sun (2022) opened the door to control spatio-temporal chaotic systems using EnKF and ensemble forecasting. Following their approach, studies such as Sun et al.

(2023), Ouyang et al. (2023), Kawasaki and Kotsuki (2024), and the present study have pursued a control simulation experiment in a similar vein. However, the proposed control algorithm in this study substantially deviates from the previous efforts. The earlier studies used EnKF solely for state estimation. Although they used ensemble forecasting derived from EnKF's analysis ensemble for control purposes, their methods for control were not directly linked to EnKF. In contrast, this study indicates that EnKF alone is sufficient for both state estimation and control. Inspired by the similarity between EnKF (or EnKS) and model predictive control, I successfully control the spatio-temporal chaotic system without altering the core architecture of EnKF.

There are several limitations of the proposed method against conventional model predictive control methods. First, the proposed method performs a control intervention only at the beginning of control horizons, which is apparently sub-optimal to minimize equation (2). Note that it is infeasible to directly optimize equation (2) for geoscientific applications such as weather modifications since iterative integrations of dynamic models are computationally expensive. I tried some heuristic methods to apply interventions at all timesteps. I divided $x_0 - \overline{x_0^a}$ obtained in Step 3 (see Algorithm 1) by the number of timesteps of the data assimilation window and applied them to model trajectory at all timesteps in the window. This method can achieve the control objective under a sufficiently small $C^r$ (not shown) mitigating the disadvantage of the proposed method. Since recent advanced observation systems enable rapid (30 seconds ~ 10 minutes) updates of data assimilation (e.g., Miyoshi et al. 2016; Sawada et al. 2019), the proposed algorithm can frequently change the strategy of interventions, which also mitigates this disadvantage of the proposed method.

Second, the proposed method cannot consider the economic cost of interventions while the original model predictive control methods can explicitly describe it in $C^u$ in Equation (2). A simple countermeasure is to scale ensembles to follow a specified variance (i.e., cost). Also, it is heuristically effective to scale an estimated control perturbation to a specified norm, which users believe is economically reasonable, as Miyoshi and Sun (2022) did.

Although EnKC outperforms the existing algorithm in the Lorenz 63 model, the potential of EnKC is not fully leveraged in the application of the low dimensional model. It is expected that EnKC is the efficient, flexible, and robust method to control large-scale phenomena as EnKF is such a method to predict Earth systems. Future works should focus on developing efficient control methods for realistic control problems such as weather modification based on EnKC.

**Acknowledgements**

This work was supported by the JST Moonshot R&D program (Grant JMPJMS2281).

| |
|---|
| **Algorithm 1**. Ensemble Kalman Control (EnKC) |
| Step 1. Perform an ETKF analysis step using forecast ensemble $x_t^{b(i)}$ and actual observations to get analysis ensemble $x_t^{a(i)}$. |
| Step 2. Compute $x_{t+T_c}^{b(i)} = M(x_t^{a(i)})$. $x_{t+T_c}^{b(i)}$ is the ensemble of extended forecast which will be used in control ETKS analysis of Step 3. |
| Step 3. Perform an ETKS analysis step using ensemble from extended forecast $x_{t+T_c}^{b(i)}$, the operator $H^c$, and a reference vector $r_{t+T_c}$ as pseudo observations. |
| Step 4. Add the perturbation $x_0 - \overline{x_0^a}$ obtained in Step 3 to the real nature. The same perturbation is also added to all analysis ensemble members at time $t$ to accurately estimate the modified nature. |
| Step 5. Compute $x_{t+T}^{b(i)} = M(x_t^{a(i)})$ to get forecast ensemble. Note that $T$ is data assimilation window length. Go back to Step 1. |

| **Algorithm 2**. Miyoshi and Sun (2022) in the Lorenz 63 model |
|---|
| Step 1. Perform an ETKF analysis step using forecast ensemble $x_t^{b(i)}$ and actual observations to get analysis ensemble $x_t^{a(i)}$. |
| Step 2. Compute $x_{t+T_c}^{b(i)} = M(x_t^{a(i)})$. $x_{t+T_c}^{b(i)}$ is the ensemble of extended forecast. If at least one ensemble member shows the regime shift ($X < 0$), the control step is activated. Otherwise, go to Step 4-2. If all the ensemble members show the regime shift, recompute the extended forecast by $x_{t+T_c}^{b(i)} = M(x_{t-T}^{a(i)})$. |
| Step 3. Pick up two ensemble members. One member shows the regime shift, and the other member does not show the regime shift. Take the difference of state variables between the two members during $t$ to $t + T$, and the obtained vectors at every timestep are scaled by the prescribed norm, $D_{fix}$. This scaled vector is used for a perturbation added to the system. Go to Step 4-1. |
| Step 4-1 (Control case). Add the perturbation obtained in Step 3 to the real nature at all timesteps within data assimilation window. When computing $x_{t+T}^{b(i)} = M(x_t^{a(i)})$ to get forecast ensemble, this perturbation is also added to all ensemble members at all timesteps within data assimilation window. Go back to Step 1. |
| Step 4-2 (No control case) Compute $x_{t+T}^{b(i)} = M(x_t^{a(i)})$ to get forecast ensemble. Go back to Step 1. |

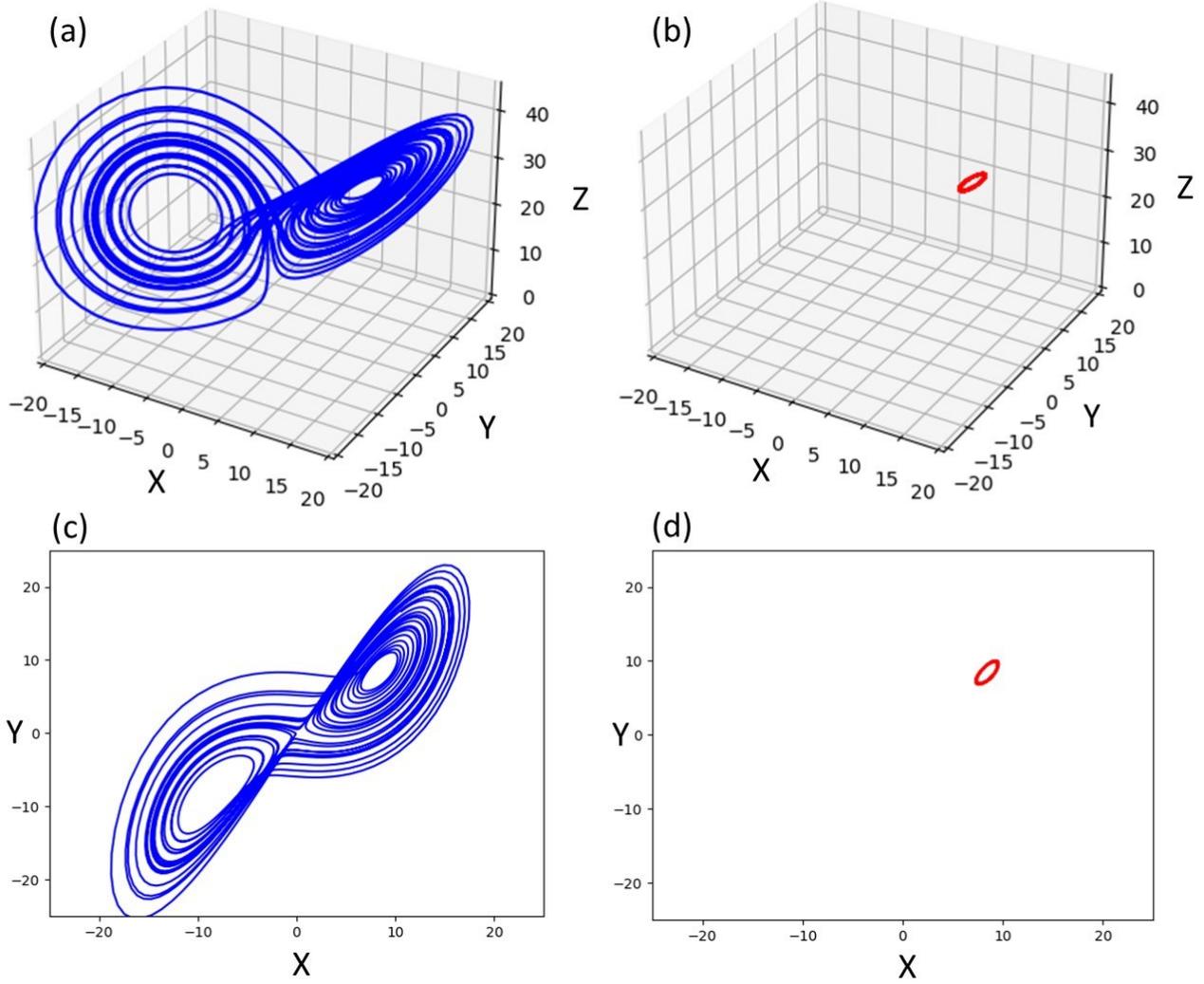

**Figure 1.** The attractor of (a) the uncontrolled Lorenz 63 nature run and (b) the controlled Lorenz 63 nature run with $C^r = 10^{-6}$ and $T_c = 300$. Both (a) and (b) shows 3000 timesteps from $5001^{st}$ timestep to $8000^{th}$ timestep. (c-d) Same as (a-b) but for the projection of a X-Y plane.

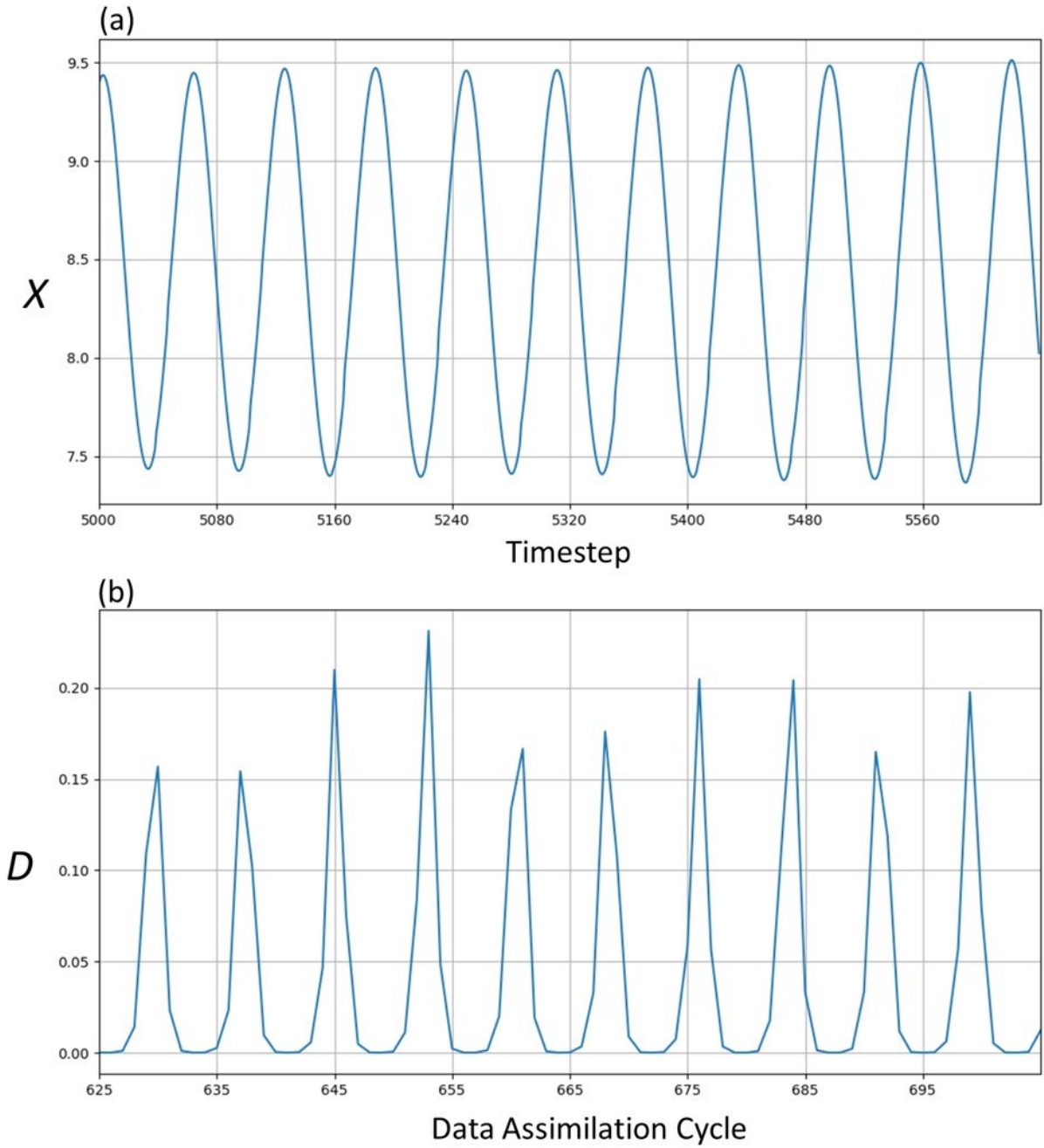

**Figure 2**. Timeseries of (a) the $X$ variable of the controlled Lorenz 63 attractor and (b) the magnitude of control perturbations, $D$, with $C^r = 10^{-6}$ and $T_c = 300$. Note that (a) and (b) show the same period.

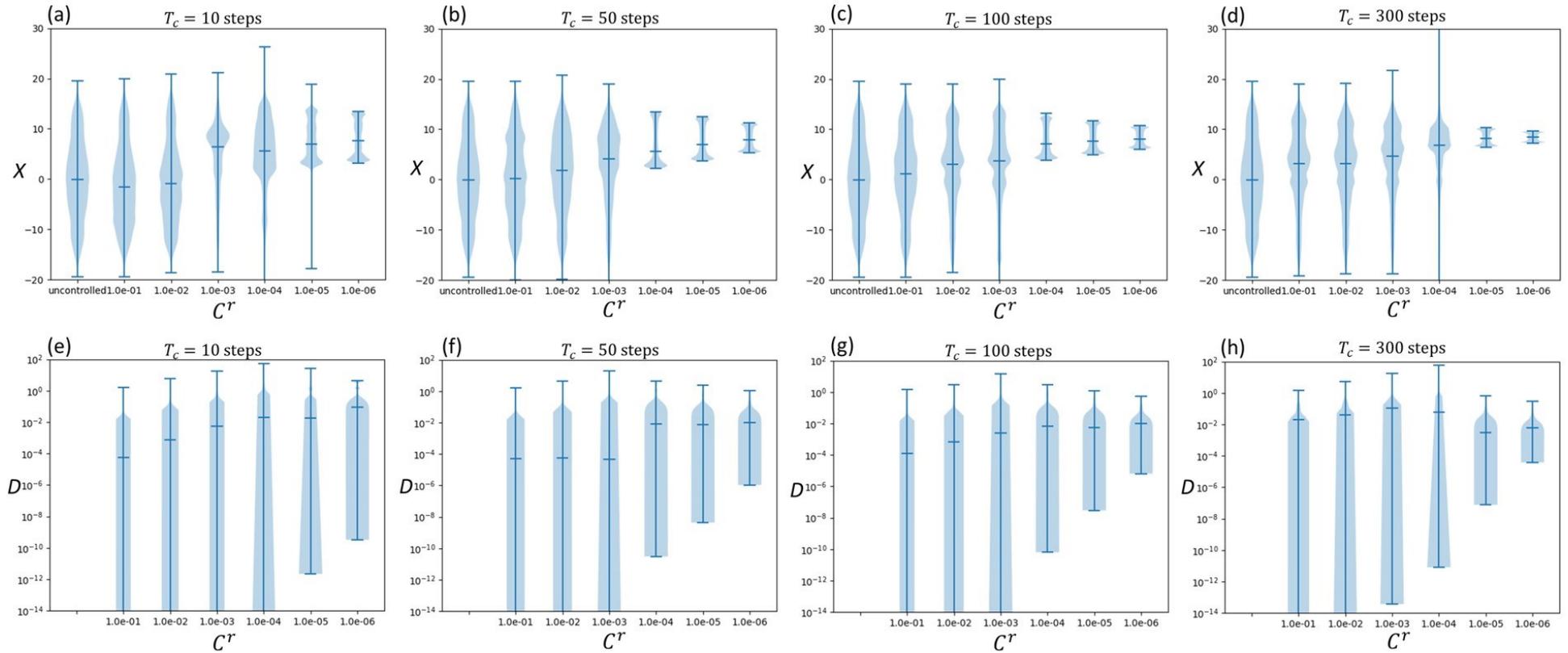

**Figure 3.** Distributions of the *X* variable of the nature during control simulation experiments with different $C^r$ under (a) $T_c = 10$, (b) $T_c = 50$, (c) $T_c = 100$, and (d) $T_c = 300$. The maximum, median, and minimum values are also shown. The leftmost violin plots show the uncontrolled experiment. (e-h) Same as (a-d) but for the magnitude of control perturbations, *D*.

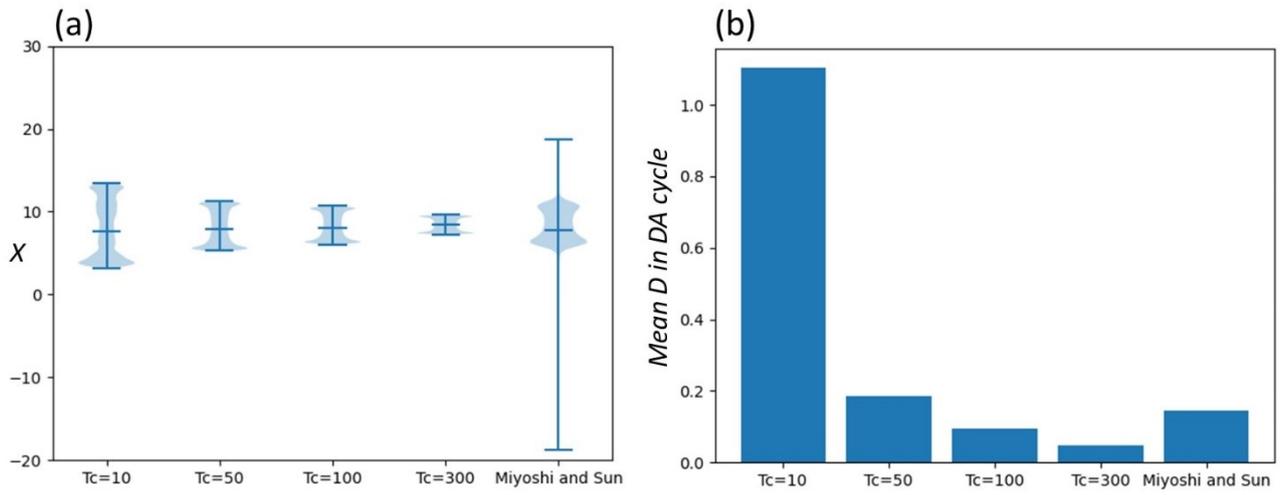

**Figure 4.** (a) Distributions of the X variable during control simulation experiments with different $T_c$ under $C^r = 10^{-6}$. The maximum, median, and minimum values are also shown. The rightmost violin plot shows the controlled experiment by the methods of Miyoshi and Sun (2022) with $T_c = 300$ and $D_{fix} = 0.05$. (b) Mean magnitude of control perturbation in the data assimilation window (8 timesteps).